\documentclass[journal=large,manuscript=article]{cup-journal}

\usepackage{amsmath,amssymb,amsfonts,latexsym,mathrsfs,nccmath}
\usepackage{bm}
\usepackage{url}
\usepackage[hidelinks]{hyperref}
\usepackage{subfig}
\usepackage{graphicx}
\usepackage{comment}
\usepackage{siunitx}
\usepackage[nopatch]{microtype}
\usepackage{booktabs}
\usepackage{braket}
\usepackage[nopatch]{microtype}
\usepackage{booktabs}
\usepackage{color}
\usepackage{notoccite}

\def\IN{\text{IN}}
\def\OUT{\text{OUT}}

\title{Robustness of Energy Landscape Controllers for Spin Rings under Coherent Excitation Transport}

\author{Sean P.\ O'Neil}
\affiliation{Department of Electrical and Computer Engineering, University of Southern California, Los Angeles, CA 90089} 
\email[S O'Neil]{seanonei@usc.edu}

\author{Frank C.\ Langbein}
\affiliation{School of Computer Science and Informatics, Cardiff University, Cardiff CF24 4AG, UK} 
\email[F Langbein]{frank@langbein.org}

\author{Edmond Jonckheere}
\affiliation{Department of Electrical and Computer Engineering, University of Southern California, Los Angeles, CA 90089}
\email[E Jonckheere]{jonckhee@usc.edu}

\author{S.\ Shermer}
\affiliation{Faculty of Science and Engineering, Physics, Singleton Park, Swansea, SA2 8PP, UK}
\email[S Shermer]{s.m.shermer@gmail.com}


\begin{document}

\begin{abstract}
The design and analysis of controllers to regulate excitation transport in quantum spin rings presents challenges in the application of classical feedback control techniques to synthesize effective control, and generates results in contradiction to the expectations of classical control theory. This paper examines the robustness of controllers designed to optimize the fidelity of an excitation transfer to uncertainty in system and control parameters. We use the logarithmic sensitivity of the fidelity error as the robustness measure, drawing on the classical control analog of the sensitivity of the tracking error. Our analysis shows that quantum systems optimized for coherent transport demonstrate significantly different correlation between error and the log-sensitivity depending on whether the controller is optimized for readout at an exact time $T$ or over a time-window $T \pm \Delta/2$. 
\end{abstract}

\section{Introduction}

Excitation transfer in the single-excitation subspace of a ring of spin-$1/2$ particles coupled via XXZ couplings forms a simple model for information transfer in a spintronic router \parencite{Schirmer2015, Schirmer2018}. Design of controls for such systems is non-trivial. Most fundamentally, measurement of the quantum state in the usual feedback control paradigm would alter the dynamics of the quantum system in a probabilistic manner \parencite{Wiseman2009}. Additionally, the coherent dynamics of a quantum system result in trajectories that evolve unitarily with all eigenvalues on the imaginary axis and are thus not asymptotically stable \parencite{Weidner2022}. Taken together, this precludes the application of common linear control techniques such as pole placement and Linear-Quadratic-Gaussian (LQG) design.

To obviate such roadblocks to development of classically-inspired controls, the work this analysis is based on appeals to the solution of a non-convex optimization problem to generate optimal, time-independent controllers \parencite{Schirmer2015}. The controllers considered are designed to alter the energy landscape of a quantum ring via static bias fields to facilitate the transfer of a single excitation from an initial spin $\ket{\IN}$ to a target spin $\ket{\OUT}$ with maximum fidelity at a given time $T$ or over a (readout) time window $T \pm \Delta/2$ under unitary dynamics. 

While design of realizable controllers is a challenge, ensuring these controllers' robustness to external perturbations or parameter uncertainty is necessary to fully harness any benefits of emerging quantum technology \parencite{Shermer2023,Glaser2015}. To progress from the current Noisy Intermediate-Scale Quantum (NISQ) era, and turn theoretical promises into reproducible experimental realities, the need for \emph{robustness} of quantum control systems emerges with accrued urgency. This is reminiscent of the situation in classical control starting nearly half a century ago, when super-maneuverable aircraft became a reality, and flight-by-wire control systems took over pilots' inputs to counter the uncertainty in the airframe model at the edge of the flight envelope---a concept that became known as \emph{robustness} and has ultimately led to the development of classical robust control. In the quantum arena, various control designs for specific applications claiming robustness have been proposed \parencite{Kosut2022, Ram2022, Zhang2022, Valahu2022, Koswara2021, Dridi2020, Wu2019, Utkan2019, Shapira2018, Deng2017, Daems2013} but a comprehensive framework for robust control for quantum systems is lacking.  

Among the ad-hoc techniques that have been developed, some have challenged \emph{physical limitations} such as the Heisenberg limit, but quantum robustness has not yet matured into 
\CUPTWOCOL
a \emph{theory of control limitations}---parallel to the very successful robustness theory developed in the 1980s for classical control systems~\parencite{Safonov_Laub_Hartmann}, which led to the formulation of quantifiable limitations on achievable performance in terms of accuracy versus sensitivity of the accuracy to uncertainties. Unfortunately, the unique characteristics of quantum systems present challenges in analyzing the robustness of control schemes in the context of classical robust control. The marginal stability of open quantum systems precludes the use of common small gain theorem-based techniques such as structured singular value analysis~\parencite{zhou:robust} in most cases. Also, in contrast to classical control problems based on asymptotic response, excitation transfer is an inherently time-domain problem requiring a time-domain view of robustness that differs from classical frequency-domain methods~\parencite{Sontag1998,oneil_2022}. 

In this analysis paper, we explore the design of time-optimal controllers published as \textcite{DataSet1} and analyze their robustness through a time-domain logarithmic sensitivity measure. The correlation between error and log-sensitivity of the controllers in this data set was first explored in \textcite{Jonckheere2018} and identified non-conventional trends for controllers optimized for time-windowed readout. In this paper we expand the analysis and include controllers optimized for instantaneous readout to better understand the robustness of the entire range of possible controllers, leading to the identification of factors that yield greater robustness. The analysis shows that controllers optimized for exact-time excitation transfer exhibit behavior in the trade-off between robustness and performance expected of a classical feedback control system. In contrast, those controllers optimized to maximize transfer over a time-window display trends between performance and robustness in contradiction to expectations fram classical control. Furthermore, in this analysis we apply a modified log-sensitivity calculation that accounts for averaging over the readout window, a factor not accounted for in previous work.

The remainder of this paper is organized in the following manner. In Section~\ref{system} we present the mathematical model for a spin-1/2 ring, derive the evolution for excitation transfer, and define the performance measure of fidelity. In Section~\ref{optimize} we present the optimization scheme for maximizing the fidelity, and in Section~\ref{robustness} we define the time-domain log-sensitivity used to gauge the robustness of the controllers. In Section~\ref{analysis} we present the hypothesis testing used to judge the conventional versus non-conventional relationship between performance (measured as the fidelity) and robustness (measured as the size of the log-sensitivity). We then present the results of the hypothesis test and identify additional robustness features not highlighted by the statistical analysis. We conclude in Section~\ref{conclude}.

\section{Methods} \label{methods}

\subsection{System Description, Dynamics, and Fidelity} \label{system}

Consider a set of $N$ interacting spin-$1/2$ particles with only one spin in an excited state and the remainder in the ground state. In this single-excitation sub-space, the network can be represented by a $N \times N$ total Hamiltonian $H_0$ with
\begin{equation} \label{eq:total}
H_0 = \hbar \sum_{m \neq n} J_{mn} (X_m X_{n}+Y_m Y_{n} +\kappa Z_m Z_{n}).
\end{equation}
Here, $J_{mn}$ are the couplings between spins $m$ and $n$, measured in units of frequency, and $\hbar$ is the reduced Planck constant.  In general $J_{mn} = J_{nm}$, and for a ring topology with nearest-neighbor coupling, $J_{mn}$ is only non-zero for $n = m \pm 1$ and $J_{1N} = J_{N1}$. In particular, we consider the case of uniform coupling, where all non-zero couplings have the same value $J$
The terms $X_n$, $Y_n$, $Z_n$ are the Pauli spin operators acting on spin $n$. These are $N$-fold tensor products whose $n$th factor is one of the Pauli matrices
\begin{equation*}
  X = \begin{pmatrix} 0 & 1 \\ 1 & 0 \end{pmatrix}, \quad
  Y = \begin{pmatrix} 0 & -\imath \\ \imath & 0 \end{pmatrix}, \quad
  Z = \begin{pmatrix} 1 & 0 \\ 0 & -1 \end{pmatrix},
\end{equation*}
and all other factors are the $2 \times 2$ identity matrix $I$. The parameter $\kappa$ distinguishes different coupling types such as XX-coupling ($\kappa=0$) or Heisenberg coupling ($\kappa=1$); specifically we consider XX-coupling. We justify this restriction to XX-coupling based on the control scheme introduced in Section~\ref{optimize}. In short, this scheme is based on spin-addressable bias fields modeled as diagonal elements of the Hamiltonian. As Heisenberg coupling introduces purely diagonal coupling terms into the Hamiltonian, they can be absorbed into the diagonal control elements so that the system model is equivalent to a strictly XX-coupled system. 

We represent the state of the system by a wavevector $\ket{\psi} \in \mathbb{C}^N$ whose $n$th entry represents the state of spin $n$. We only consider normalized wavevectors such that $\left|\braket{\psi|\psi}\right|^2 = 1$. Specifically, if spin $n$ is measured to be in the excited state with absolute certainty, the $n$th entry of $\ket{\psi}$ has magnitude $1$. Conversely, if the spin has zero probability of being excited the entry is $0$, indicating the spin is in the ground state. A value of $0<|\psi_n|<1$ indicates the $n$th spin has a non-zero probability to be excited. If the state $\ket{\psi}$ differs from the state $\ket{\psi_0}$ only by a phase factor $e^{\imath \varphi}$ then $\left|\braket{\psi_o|\psi}\right|^2 = 1$. Associating the $N$ state vectors $\{\ket{\psi_n}\}$ which indicate a single excitation on spin $n$ with the natural basis vectors of $\mathbb{C}^N$ provides a convenient basis for describing the system dynamics. 

In this basis, considering only XX-coupling and ring topology, the Hamiltonian of~\eqref{eq:total} takes the explicit form
\begin{equation}\label{e:Hamiltonian}
H_0 = \hbar\begin{pmatrix}
  0      & J & 0      & \ldots & 0      & J\\
  J      & 0 & J      &        & 0      & 0\\
  0      & J & 0      &        & 0      & 0\\
  \vdots &   & \ddots & \ddots & \ddots & \vdots\\
  0      & 0 & 0      &        & 0      & J\\
  J      & 0 & 0      & \ldots & J      & 0
\end{pmatrix}.
\end{equation}
The dynamical evolution of this system is governed by the time-dependent Schr\"odinger equation:
\begin{equation} \label{eq: schrodinger}
\frac{d}{dt} \Ket{\psi(t)} = -\frac{\imath}{\hbar} H_0 \Ket{\psi(t)}, \quad \Ket{\psi(0)} = \Ket{\psi_0}.
\end{equation}
Assuming a system of units where $\hbar = 1$, the solution to~\eqref{eq: schrodinger} is 
\begin{equation} 
\Ket{\psi(t)} = e^{-\imath t H_0} \Ket{\psi_0} = U(t) \Ket{\psi_0}.
\end{equation}

Noting that $H_0$ is Hermitian with real eigenvalues, we can immediately see that the eigenvalues of the open-loop system are purely imaginary, and so the system is not stable, but only marginally stable \parencite{Chen2013}. In simplest terms, this means there is no asymptotic steady state of the system, as evident from the eigenvalues of the form $\{-\imath\lambda_n\}_{n=1}^{N}$. This presents two conflicting issues in the control of closed quantum systems. On the one hand, unitary evolution of the system is desirable in retaining the coherence or phase of the system, which is a key feature that gives quantum technology an advantage over classical technologies. On the other hand, the techniques of classical control theory (pole-placement, LQG, etc.) require synthesis of \emph{stabilizing} controllers \parencite{dorf}. While this is prudent from a classical point of view in that stabilizing controllers preclude the possibility of an unbounded response, applied to a quantum system, this would result in convergence to a classical steady state, resulting in the loss of coherence. This provides a strong motivation for the development of control techniques outside the scope of established classical feedback control.

We now consider the problem of transferring the single excitation of the system from a given input spin $\ket{\IN} = \ket{\psi_0}$ to a specific output spin $\ket{\OUT}$. At a given time $T$ the probability that $\ket{\psi(T)} = \ket{\OUT}$ modulo the global phase $e^{\imath \varphi}$ is given by the squared overlap of the current state with the target state or 
\begin{equation} \label{eq: fidelity1}
  \mathcal{F}(T) = \left|\Braket{\OUT|\psi(T)}\right|^2 = \left|\Braket{\OUT |U(T)|\IN}\right|^2,
\end{equation}
where $\mathcal{F}(T)$ is the \emph{fidelity} of the transfer at time $T$. Extending this concept to a time-window of $\pm \Delta/2$ about the time $T$, we define the time-averaged fidelity as 
\begin{equation} \label{eq:fidelity2}
  \mathcal{F}(T \pm \Delta/2) = \frac{1}{\Delta } \int_{T - \Delta/2}^{T +\Delta/2} \left|\Braket{\OUT|U(t)|\IN}\right|^2 dt.
\end{equation}
Finally, noting that the upper bound on both $\mathcal{F}(T)$ and $\mathcal{F}(T \pm \Delta/2)$ is unity, we define the fidelity error in analogy to the tracking error as 
\begin{equation}
\begin{aligned} \label{eq:error}
e(T) &= 1 - \mathcal{F}(T),\\
e_{\Delta}(T) &= 1 - \mathcal{F}(T \pm \Delta/2).
\end{aligned}
\end{equation}

\subsection{Design Goals and Optimization Scheme}\label{optimize}

Consider the design goal of maximizing the fidelity for the instant time case~\eqref{eq: fidelity1} or the time-averaged case~\eqref{eq:fidelity2}. To obviate the issues of backaction involved in measurement-based feedback control we introduce control via static bias-fields that ideally address a single spin to alter the energy landscape of the system. In terms of the Hamiltonian~\eqref{e:Hamiltonian}, these control fields take the form
\begin{equation}
   D = \sum_{n=1}^N D_{n} = \sum_{n=1}^N d_n \ket{n}\bra{n}.
\end{equation}
Here, the $D_n \in \mathbb{R}^{N \times N}$ consist of all zeros, save for the $n$th diagonal element which assumes the scalar value $d_n$ of the field addressing spin $n$. This augments the natural Hamiltonian so that $H_D = H_0 + D$. The state transition matrix is thus modified as $U_D(t) = e^{-\imath t(H_0 + D)}$, and the expressions for the fidelity in~\eqref{eq: fidelity1} and~\eqref{eq:fidelity2} are similarly modified. 

Maximization of the fidelity at a specific time $T$ or over a window $T \pm \Delta/2$ then becomes a non-convex optimization problem of the form 
\begin{equation} \label{eq:optimize1}
\min_{\{D_n,T\} \in \mathbb{X}} \left[ 1 - \left| \bra{\OUT}e^{-\imath T(H_0 + \sum_n D_n)}\ket{\IN} \right|^2 \right]
\end{equation}
or
\begin{equation} \label{eq:optimize2}
\min_{ \{D_n,T\} \in \mathbb{X}} \left[1 - \frac{1}{\Delta} \int_{T - \Delta/2}^{T + \Delta/2} \left| \bra{\OUT}e^{-\imath t(H_0 + \sum_n D_n)}\ket{\IN} \right|^2 dt \right]. 
\end{equation} 
Here, $\mathbb{X}$ defines the set of admissible controllers $D_n$ and readout times $T$ defined by the optimization constraints.

The controllers used in this study were developed using the MATLAB's \texttt{fminunc} solver with the BFGS quasi-Newton algorithm. The optimization was performed with a bias toward producing high fidelity controllers by choosing start times corresponding to high-fidelity peaks in the transfer of an equivalent chain between $\ket{\IN}$ and $\ket{\OUT}$ as initial values for the time variable. Furthermore, we placed symmetry conditions on the possible values of $D_n$. Specifically, the $D_n = d_n \ket{n} \bra{n}$ were constrained so that $d_{\IN} = d_{\OUT}$ and $d_{\IN + k} = d_{\OUT - k}$ for $k \in \{1 \hdots \lceil (\OUT-\IN)/2 \rceil \}$. See \textcite{Schirmer2015} for a more detailed exposition of the optimization and constraints.

\subsection{Robustness Measure---Log-Sensitivity}\label{robustness}

Given a system model and controls to maximize the fidelity, we consider the issue of robustness of the control scheme to uncertainty in the system parameters or control fields. We denote an uncertain parameter (coupling coefficient or bias field) as $\xi_{\mu} \in \mathbb{R}$ such that
\begin{equation} \label{xi}
  \xi_{\mu} = 
    \begin{cases}
    d_{\mu} + \delta_{\mu}, & 1 \leq \mu \leq N,\\
    J_{(\mu-N),(\mu-N+1)} + \delta_{\mu}, & N+1 \leq \mu \leq 2N-1,\\
    J_{1,N} +\delta_\mu, & \mu = 2N,
    \end{cases}
\end{equation}
so that $\mu \in \{1, \hdots, N\}$ correspond to perturbations to the control and $\mu \in \{N+1, \hdots , 2N\}$ correspond to perturbations to the Hamiltonian. Here, $\delta_\mu \in \mathbb{R}$ represents the deviation from the nominal value in compatible physical units with $\xi_\mu$.

These uncertainties enter the Hamiltonian through structure matrices $S_{\mu} \in \mathbb{R}^{N \times N}$. The uncertain Hamiltonian becomes $\tilde{H}_D = H_0 + D + \sum_{\mu} \delta_{\mu} S_{\mu}$. Specifically we define 
\begin{equation} \label{eq:Smu}
  S_{\mu} = \begin{cases}
    \ket{\mu}\bra{\mu}, &1 \leq \mu \leq N,\\
    \ket{\mu-N}\bra{\mu-N+1} + \\ 
      \quad \ket{\mu-N+1}\bra{\mu-N}, &N+1 \leq \mu \leq 2N-1,\\
    \ket{1}\bra{N} + \ket{N}\bra{1}, & \mu = 2N.
\end{cases}
\end{equation}
Consequently, we have the uncertain state-transition matrix as $\tilde{U}(t) = e^{-\imath t(H_0 + D + \sum_{\mu} \delta_{\mu} S_{\mu})}$. Considering a single uncertain parameter in the Hamiltonian, we look at the differential sensitivity of the state transition matrix to that parameter as
\begin{align} \label{eq:partial}
\begin{split}
  &\frac{\partial \tilde{U}(T)}{\partial \xi_{\mu}} = \lim_{ \xi_{\mu} \to \xi_{\mu0}} \frac{e^{-\imath t\tilde{H}_D(\xi_{\mu})} -e^{-\imath t\tilde{H}_{D}(\xi_{\mu0})}}{\xi_{\mu} - \xi_{\mu0}},\\
  &\frac{\partial \tilde{U}(T)}{\partial \delta_{\mu}}=\lim_{\delta_{\mu} \to 0} \frac{e^{-\imath t(H_0 + D + \delta_{\mu} S_{\mu})} -e^{-\imath t(H_0 + D) }}{\delta_{\mu}},
\end{split}
\end{align}
where $\xi_{\mu}$ is defined as in~\eqref{xi} with nominal value given by $\xi_{\mu0}$ when $\delta_{\mu} = 0$. 

We note the differential sensitivity of~\eqref{eq:partial}, in both equivalent forms, is valuable in its own right to measure the effect of parameter uncertainty on $e(T)$. However, this pure differential sensitivity carries an intrinsic scaling by the physical units of the parameter in the denominator of the limit. While this permits a useful comparison in sensitivity for the same type of uncertainty, it does not provide an unbiased measure for comparing robustness between different uncertainty categories. For this reason, we seek a \emph{dimensionless} measure of robustness in the logarithmic sensitivity, requiring renormalization of the terms in~\eqref{eq:partial} by $U^{\dagger}(T)\xi_{\mu0}$ or $U^{\dagger}(T) \delta_{\mu0}$. Even though $\partial/\partial \xi_\mu=\partial/\partial \delta_u$, these two normalization factors result in different log-sensitivities. This is obvious by noting that $U^\dagger(T)\xi_{\mu0}\ne 0$ while $U^\dagger(T)\delta_{\mu0}=0$. Finally, observe that if the uncertain parameter has a nominal value of zero, the log-sensitivity formulation above requires modification to consider only deviations from the nominal value while producing a non-trivial measure of sensitivity. 

Noting that the performance measure $\mathcal{F}(\cdot)$ is time-based, we assess the robustness of the control scheme by determining the differential effect of uncertainty on the fidelity error $e(T)$ or $e_{\Delta}(T)$ as defined in~\eqref{eq:error} (equally the fidelity) for instantaneous readout as 
\begin{equation} \label{eq:log-sens}
s(\xi_{\mu0},T) = \left. \frac{\partial e(T)}{\partial \xi_{\mu}} \frac{\xi_{\mu}}{e(T)} \right|_{\xi_{\mu0}}
\end{equation}
and for time-windowed readout as 
\begin{equation}
s_{\Delta}(\xi_{\mu0},T) = \left. \frac{\partial e_{\Delta}(T)}{\partial \xi_{\mu}} \frac{\xi_{\mu}}{e_{\Delta}(T)} \right|_{\xi_{\mu0}}.
\end{equation}
We see that~\eqref{eq:log-sens} is the differential sensitivity of the fidelity error normalized by the ratio of the nominal parameter value and nominal fidelity error.

Consider a decomposition of $e^{-\imath t(H_0 + D)} = \sum\limits_{n = 1}^N \Pi_n e^{-\imath t\lambda_n}$, and let $\omega_{mn} = \lambda_m - \lambda_n$. Here $\Pi_n$ are the projectors onto the orthogonal subspaces of the controlled Hamiltonian $H_D$. Specifically, from the spectral decomposition of $H_D = V \Lambda V^{\dagger}$, $\lambda_n$ is the $n$th diagonal entry of $\Lambda$ and $\Pi_n$ is the dyadic product of the $n$th column of $V$ with itself or  $\Pi_n = V_n V_n^{\dagger}$. Then, for $e(T) = 1 - \mathcal{F}(T)$, we have from \textcite{Schirmer2018}
\begin{align} \label{eq:sens_t}
\frac{\partial{e(T)}}{\partial \delta_{\mu}}& = 2T \sum_{m,n}\bra{\OUT} \Pi_{m} S_{\mu} \Pi_{n} \ket{\IN}\operatorname{sinc} \left(\tfrac{1}{2}T\left(\omega_{mn} \right) \right) \nonumber\\ 
&\times \sum_{p} \bra{\IN} \Pi_{p} \ket{\OUT} \sin\left( \tfrac{1}{2} T \left( \omega_{mp} + \omega_{np} \right) \right),
\end{align}
where $\operatorname{sinc}(x) = \frac{\sin(x)}{x}$. For $e_{\Delta}(T) = 1-\mathcal{F}(T \pm \Delta T)$ we have a more complicated expression,
\begin{equation} \label{eq:sens_dt}
\frac{\partial e_{\Delta}(T)}{\partial \delta_{\mu}} = \sum_{\lambda_m = \lambda_n \neq \lambda_p} A(T,\Delta,\lambda_m, \lambda_p) + \sum_{\lambda_m \neq \lambda_n}B(T,\Delta,\lambda_m,\lambda_n,\lambda_p).
\end{equation}
Note that for $\lambda_n = \lambda_m = \lambda_p$ there is no contribution to the sum. Specifically we have
\begin{multline}
A(T,\Delta,\lambda_m,\lambda_p) = \\
\frac{1}{\Delta} \bra{\OUT} \Pi_p \ket{\IN} \bra{\IN} \Pi_m S_{\mu} \Pi_{m} \ket{\OUT} \times \\ 
\left\{\tfrac{2}{\omega_{mp}} \left[ (T+\tfrac{\Delta}{2})\cos[\omega_{mp}(T+\tfrac{\Delta}{2})] \qquad\right.\right.\\
  \qquad\left.- (T-\tfrac{\Delta}{2})\cos[\omega_{mp}(T-\tfrac{\Delta}{2})] \right] \\ 
- \  \tfrac{2}{\omega_{mp}^2}  \left(\sin[\omega_{mp}(T+\tfrac{\Delta}{2})] - \sin[\omega_{mp}(T - \tfrac{\Delta}{2})] \right) \Big\}
\end{multline}
and 
\begin{multline}
B(T,\Delta,\lambda_m,\lambda_n,\lambda_p) = \\
 \frac{1}{\Delta} \frac{2}{\omega_{mn}}\bra{\OUT}\Pi_p\ket{\IN} \bra{\IN}\Pi_m S_{\mu} \Pi_{n} \ket{\OUT}  \\ 
 \times \left( (T+\tfrac{\Delta}{2})\operatorname{sinc} \left[ \omega_{mp} (T+\tfrac{\Delta}{2}) \right] -   (T-\tfrac{\Delta}{2})\text{sinc}\left[ \omega_{mp} (T-\tfrac{\Delta}{2}) \right] \right.\\
 - (T+\tfrac{\Delta}{2})\operatorname{sinc}\left[ (\omega_{np}) (T+\tfrac{\Delta}{2}) \right]\\
 \left. + (T-\tfrac{\Delta}{2})\operatorname{sinc}\left[ (\omega_{np} (T-\tfrac{\Delta}{2}) \right] \right).
\end{multline}
We use the differential sensitivity established by~\eqref{eq:sens_t} and~\eqref{eq:sens_dt}, normalized by the ratio $\frac{\xi_{\mu0}}{e(T)}$ or $\frac{\xi_{\mu0}}{e_{\Delta}(T)}$, to get a non-trivial, i.e., non-vanishing,  log-sensitivity as robustness measure.

\section{Analysis}\label{analysis}

Our analysis of the controllers produced by the optimization in Section~\ref{optimize} consists of two parts: (1) statistical hypothesis testing of the relationship between performance, as measured by the size of the fidelity error, and robustness, gauged by the size of the log-sensitivity and (2) identification of areas that require more exploration to explain the observed robustness properties.

\subsection{Classical Control Considerations}
 
To relate to classical robustness as constructed in the 1980s, and motivate the hypothesis tests performed, we compare the problem of state transfer $\ket{\IN} \to \ket{\OUT}$ under a Hamiltonian $H_D$ containing the control terms considered in this paper, itself a paradigmatic problem in quantum control, with the classic paradigmatic problem of transfer of the zero input state $0$ to a constant output state $1$ in the simple Single Input Single Output (SISO) control, chosen for ease of the exposition.

The accuracy of such transfer, be it quantum or classical, can be formulated in terms of a \emph{sensitivity} operator that maps the desired output to the tracking error, here defined as the difference between the desired output and the actual output, 
\begin{equation}\label{eq:classicalS}
    \varepsilon(t)=1(t)-\int_0^t\!\!\! T(t-\tau)1(\tau)d\tau
    = \int_0^t\!\!\! [\delta(t-\tau)-T(t-\tau)]1(\tau)d\tau,
\end{equation} 
where $1(t)$ is the Heaviside unit step, $\delta(t)$ the Dirac delta, and $T(t)$ is the impulse response of the control system from the desired output to the actual output. Rewriting the above relationship as $\varepsilon(t)=\int_0^tS(t-\tau)1(\tau)\,d\tau$ defines the \textbf{\emph{sensitivity operator}} $S$, quantifying accuracy. Since the seminal work of \parencite{Bode1945} motivated by feedback amplifiers, classical control has formulated the limitations in terms of the \emph{Laplace transforms} of the operators, $\widehat{S}(s)$ and $\widehat{T}(s)$. Elementary manipulation in the Laplace domain reveals that $\widehat{T}(s)$ is the log-sensitivity of $\widehat{S}(s)$ relative to unstructured perturbations, and the operators satisfy the fundamental limitation 
\begin{equation}\label{eq:classical_limitation}
\widehat{S}(s)+\widehat{T}(s)=1,
\end{equation}  
forbidding simultaneous near zero error and near zero sensitivity. Only recently \parencite{oneil_2022} has this Laplace domain limitation begun to be understood in the time-domain, which is essential for quantum problems where readouts happen at a specific time.

The quantum transfer error can be formulated similarly. However, when dealing with state transfer problems involving wavefunctions or pure states (as we do here), there is an additional complication from a tracking error point of view, in that such quantum states as $\ket{\IN}$ and $\ket{\OUT}$ are defined only up to a global phase factor $\exp(\imath \varphi)$. This means that the true tracking error is the \emph{projective} error $\varepsilon_{\mathrm{proj}}(t)=\ket{\OUT}-\exp(\imath \varphi) \exp(-\imath H_D t)\ket{\IN}$. Defining the input/output swapping operator $W$ by $\ket{\IN}=W\ket{\OUT}$, the preceding can be rewritten as 
\begin{equation}\label{eq:pseudoquantumS}
  \varepsilon_\mathrm{proj}(t)
  = (I-\exp(\imath \varphi)\exp(-\imath H_D t)W)\ket{\OUT}.
\end{equation}
Except for the phase factor, the connection between~\eqref{eq:classicalS} and~\eqref{eq:pseudoquantumS} is obvious.  The phase is used to bring the error below the classical limitation by defining   
\[
  \varphi^*(t)=\arg \min_\varphi \Vert \varepsilon_\mathrm{proj}(t) \Vert =-\angle\bra{\OUT}\exp(-\imath H_D t)\ket{\IN}.
\]
Elementary complex analysis reveals that $\Vert \varepsilon_\mathrm{proj}^*\Vert^2=2(1-F)$, where $F:=|\bra{\OUT}\exp(-\imath H_D t)\ket{\IN}|$ is the \emph{overlap} between desired and actual states rather than the fidelity $\mathcal{F}=F^2$. Nevertheless, for very high fidelity, $\Vert \varepsilon_\mathrm{proj}^*\Vert^2\approx (1-\mathcal{F})$. The sensitivity of the latter will be our major concern. The difficulties in the analysis arising from the global phase can also be avoided by formulating state transfer problems in the density operator formalism.

Rewriting~\eqref{eq:pseudoquantumS} as $\varepsilon_\mathrm{proj}^*(t)=S_\mathrm{proj}(t)\ket{\OUT}$, where $S_\mathrm{proj}(t)$ is the \textbf{\emph{projective or quantum sensitivity function}}, it follows that, at the limit $\mathcal{F}\uparrow 1$,   
\begin{equation}\label{eq:e_versus_S}
  1-\mathcal{F}
  =\bra{\OUT}S_\mathrm{proj}(t)^\dagger S_\mathrm{proj}(t)\ket{\OUT}.
\end{equation}

Moreover, from~\eqref{eq:e_versus_S} and $\mathcal{F}=|\bra{\OUT}T(t)\rangle|^2$, 
where $\ket{T(t)}=\exp(-\imath H_Dt)\ket{\IN}$ is the closed-loop transfer impulse response, it follows that 
\begin{equation}\label{eq:quantum_limitation}
  \bra{\OUT}(T(t) T(t)^\dagger+S_\mathrm{proj}(t)^\dagger S_\mathrm{proj}(t)) \ket {\OUT}=1.
\end{equation}
To make the above a \emph{limitation}, let us remove the phase factor in $S_{\mathrm{proj}}(t)$, in which case it is easily seen that  
$d S_{\varphi=0}(t)\ket{\OUT}=\imath t T(t) dH_D$ for $[H_D,dH_D]=0$. In other words, as in classical control, $T(t)$ is the sensitivity of the sensitivity function.

The connection between the classical~\eqref{eq:classical_limitation} and the quantum limitation~\eqref{eq:quantum_limitation} is obvious,  but it indicates that this limitation is still classical.  However, incorporating the phase factor in the sensitivity function, as done in \textcite{CDC_phase}, could alleviate it.

\subsection{Hypothesis Test}

We establish the following two-tailed hypothesis test to confirm or refute whether the controllers in our data set \parencite{DataSet1} conform to the conventional limitations on robustness and performance established above. For brevity in the following section we describe the hypothesis testing in terms of $e(T)$ versus $s(\xi_{\mu0},T)$, but the conditions apply equally to $e_{\Delta}(T)$ and $s_{\Delta}(\xi_{\mu 0},T)$: 
\begin{itemize}
\item $H_0$: null hypothesis postulating no trend between $s(\xi_{\mu0},T)$ and $e(T)$,
\item $H_{1+}$: alternative hypothesis one postulating positive correlation between $s(\xi_0,T)$ and $e(T)$ and indicative of controllers that do not exhibit the conventional limitation on performance and robustness,
\item $H_{1-}$: alternative hypothesis two postulating negative correlation between $s(\xi_0,T)$ and $e(T)$ and indicative of controllers that exhibit the conventional limitation on performance and robustness. 
\end{itemize}

To execute the test we chose two distinct correlation measures: the Kendall $\tau$ as a non-parametric test based  on rank-ordering of the data \parencite{kendall_tau} and the Pearson $r$ linear correlation coefficient to test the linear relation between the two metrics on a log-log scale. We chose ring sizes from $N=3$ to $N=20$. For all controllers examined, the initial state is taken as $\ket{\IN} = \ket{1}$ so that the excitation is initially located at spin $1$. For the time-windowed readout controllers (the dt controllers) we tested excitation transfer ranging from localization at the initial spin $\ket{\IN} = \ket{\OUT} = \ket{1}$ up to $\ket{\OUT} = \left|\left\lceil \frac{N}{2} \right\rceil\right\rangle$. For the instant readout case (the t controllers) we consider transfers from $\ket{\OUT} = \ket{2}$ through $\ket{\OUT} = \left|\left\lceil \frac{N}{2} \right\rceil\right\rangle$. We note that there is nothing unique in the selection of $\ket{1}$ as the initial spin as the ring is rotationally symmetric. Likewise, consideration of transfers only up to $\left\lceil \frac{N}{2} \right\rceil$ is justified by the symmetry of the ring as well. This provides a total of $90$ test cases for the instantaneous readout controllers and $108$ test cases for the time-windowed readout case. Though a complete set of $2000$ controllers exists for each possible transfer, we exclude controllers that yield a fidelity $\mathcal{F} < 0.9$, and base our analysis on the remaining controllers, maintaining consistency with the analysis in \textcite{Jonckheere2018}.

To compute the degree of correlation between $e(T)$ and $s(\xi_0,T)$ for each ring and transfer combination, we apply the \texttt{corr($\cdot$,$\cdot$)} function from MATLAB with the \texttt{`Kendall'} option to produce the Kendall $\tau$ and \texttt{`Pearson'} to generate the Pearson $r$. With the raw Kendall $\tau$ and Pearson $r$ we establish the threshold for statistical significance at $\alpha = 0.01$ to reject $H_0$ in favor of $H_{1+}$ for a positive (rank) correlation coefficient and in favor of $H_{1-}$ for a negative (rank) correlation coefficient. We judge the level of significance for each possible test case depending on the correlation coefficient used. For the Kendall $\tau$ we normalize by the standard deviation so that 
$Z_{\tau} = \tau \left(\sqrt{\frac{2(2n+5)}{9n(n-1)}}\right)^{-1}$ where $n$ is the number of samples (controllers) within the test case. We then quantify the statistical significance of the results through their $p$-values defined as
\begin{equation}
  p_{\tau} = 
    \begin{cases}
    \Phi(Z_{\tau}) , & \tau < 0, \\
    1-\Phi(Z_{\tau}), & \tau > 0,
    \end{cases}
\end{equation}
where $\Phi$ is the normal cumulative distribution function. To evaluate the statistical significance of the Pearson $r$, we translate the raw correlation coefficient to a t-statistic through $t_{r} = r \left(\sqrt{\frac{1-r^{2}}{n-2}}\right)^{-1}$. We then quantify the statistical significance of the test for a given value of $r$ as 
\begin{equation}
  p_{r} = 
    \begin{cases}
    \mathcal{S}(t_{r}) , & r < 0, \\
    1-\mathcal{S}(t_r), & r > 0,
    \end{cases}
\end{equation}
where $\mathcal{S}$ represents the cumulative Student's $t$-distribution. 

Finally, though we are generally looking at the trend of $s(\xi_{\mu0},T)$ versus $e(T)$, there are a total of $2N$ perturbation directions to examine for each excitation transfer. To streamline the analysis, we focus specifically on three categories of perturbation within each possible transfer and ring: 
\begin{itemize}
\item Norm over the $N$ controller perturbations---in this case we examine the trend of $e(T)$ versus $\lVert  s(\xi_{\mu0},T) \rVert_{C} = $ \linebreak $\sqrt{\sum\limits_{\mu=1}^{N} \left| s(\xi_{\mu0},T) \right|^2}$.
\item Norm over the $N$ Hamiltonian uncertainties---in this case we examine the trend of $e(T)$ versus 
\linebreak $\lVert  s(\xi_{\mu0},T)  \rVert_{H} = \sqrt{\sum\limits_{\mu=N+1}^{2N} \left| s(\xi_{\mu0},T) \right|^2} $.
\item Norm of all $2N$ uncertainties---in this case we examine the trend of $e(T)$ versus $\lVert  s(\xi_{\mu0},T)  \rVert = \sqrt{\sum\limits_{\mu=1}^{2N} \left| s(\xi_{\mu0},T) \right|^2}$.
\end{itemize}
We present the results in the following section in terms of these uncertainty categories.

\subsection{Hypothesis Test Results}\label{results}

The entire spreadsheet depicting the results of hypothesis test is available in the repository \textcite{static_bias_controlelers}. We present the following summary of significant deductions from the hypothesis test.

\subsubsection{Instant Readout Controllers (t Controllers)}

The trend between $e(T)$ and each normed measure of $s(\xi_{\mu0},T)$, measured by the Kendall $\tau$ for rank correlation, is overwhelmingly conventional, showing a negative correlation between error and log-sensitivity, save for the transfer from spin $1$ to spin $2$ or nearest-neighbor transfer. For nearest-neighbor transfer the hypothesis test rejects $H_0$ in favor of $H_{1+}$ for all nearest-neighbor transfers for $N\geq 7$. None of the tests for the nearest-neighbor transfer fail to meet the $\alpha = 0.01$ threshold and are thus considered reliable. Though not the complete list of results, Table~\ref{table1} provides a snapshot of the hypothesis test for the correlation between $e(T)$ and $\lVert s(\xi_{\mu0},T) \rVert$ for $N=3$ to $N=12$. In detail: 
\begin{itemize}
\item For the $e(T)$ versus $\lVert s(\xi_{\mu0},T) \rVert$ correlation, five of the $90$ tests fail to achieve a significance level of $\alpha = 0.01$ and are excluded. Of the remaining tests, all display a conventional negative trend, save for the nearest-neighbor transfers noted above. 
\item Of the $90$ tests for $e(T)$ versus $\lVert s(\xi_{\mu0},T) \rVert_{C}$, all but nine display a conventional trend with a confidence of at least $99 \%$. Of these nine tests, all fall into the category of nearest-neighbor transfer, seven display a $p$-value greater than $\alpha$ and are discarded, and the other two display a non-conventional positive trend. 
\item The tests for $e(T)$ versus $\lVert s(\xi_{\mu0},T) \rVert_{H}$ follows the same pattern as that of $\lVert s(\xi_{\mu0},T) \rVert$ with nearest-neighbor transfers displaying a non-conventional trend with high confidence, except for $N < 6$ cases. Of the remaining tests, all show a conventional trend except for nine cases that fail to meet the required confidence level.
\end{itemize}

As a check on consistency, we compare the hypothesis test results based on the rank-correlation of the Kendall $\tau$ with the results based on the linear correlation coefficient of Pearson's $r$. Though not identical, the hypothesis tests based on each measure show strong agreement as summarized below:

\begin{itemize}

\item For the $\lVert s(\xi_{\mu0},T) \rVert$ tests, the hypothesis tests provide identical results in terms of acceptance or rejection of $H_0$ with two exceptions, neither of which affect the non-convention\-al trend for nearest-neighbor transfer. For the Pearson $r$-based test, the $N=6$, $1\rightarrow2$ test does not display the confidence to reject the null-hypothesis as in the Kendall $\tau$-based test. Conversely, while the $N=12,\;1\rightarrow6$ transfer is unable to reject $H_0$ for the Kendall $\tau$ test, the Pearson $r$ test does reject the null hypothesis in favor of $H_{0-}$. 

\item The comparison for $\lVert s(\xi_{\mu0},T) \rVert_{C}$ shows strong consistency, agreeing in rejection of $H_0$ in favor of $H_{1-}$ for all transfers except for $N\geq11$ nearest-neighbor transfers with one exception---the Pearson $r$ test is inconclusive for the $N=10$ nearest-neighbor transfer. Of the remaining nine nearest-neighbor tests, the Pearson $r$ test provides higher confidence, with seven of the nine rejecting $H_0$ in favor of $H_{1+}$ with high confidence. 

\item The Pearson $r$-based hypothesis test for $e(T)$ versus \linebreak $\lVert s(\xi_{\mu0},T) \rVert_{H}$ agrees with the Kendall $\tau$ in rejection of $H_0$ for all nearest-neighbor transfer for $N \geq 6$ but displays ten other cases with failure to reject $H_0$ compared to nine for the Kendall $\tau$ test.

\end{itemize}

\begin{table}[!t]
    \centering
    \caption{Excerpt of hypothesis test results for $e(T)$ versus $ \lVert s(\xi_{\mu0},T) \rVert$ using Kendall $\tau$. Note the positive trend for nearest-neighbor transfers starting with $N=7$. Also note the strong significance of the test with only the $N=12$, $1 \rightarrow 6$ transfer failing to meet the $p < \alpha = 0.01$ threshold.}
    \begin{tabular}{|c|c|c|c|}
    \hline
        Transfer & $\tau$ for $e(T)$ vs. $\lVert s(\xi_{\mu0},T) \rVert$ & $Z_{\tau}$& $p$ \\ \hline
        N=3 out=2 & -0.0512 & -3.4191 & 0.0003 \\ \hline
        N=4 out=2 & -0.1560 & -7.2404 & 0.0000 \\ \hline
        N=5 out=2 & -0.4969 & -32.5270 & 0.0000 \\ \hline
        N=5 out=3 & -0.2300 & -13.7108 & 0.0000 \\ \hline
        N=6 out=2 & -0.0436 & -2.5253 & 0.0058 \\ \hline
        N=6 out=3 & -0.6051 & -37.9070 & 0.0000 \\ \hline
        N=7 out=2 & 0.0688 & 4.1724 & 0.0000 \\ \hline
        N=7 out=3 & -0.5134 & -30.9641 & 0.0000 \\ \hline
        N=7 out=4 & -0.3464 & -19.0509 & 0.0000 \\ \hline
        N=8 out=2 & 0.0723 & 4.0931 & 0.0000 \\ \hline
        N=8 out=3 & -0.5216 & -32.5941 & 0.0000 \\ \hline
        N=8 out=4 & -0.2665 & -9.4653 & 0.0000 \\ \hline
        N=9 out=2 & 0.0757 & 4.7660 & 0.0000 \\ \hline
        N=9 out=3 & -0.4376 & -27.2378 & 0.0000 \\ \hline
        N=9 out=4 & -0.4369 & -19.8395 & 0.0000 \\ \hline
        N=9 out=5 & -0.2564 & -10.5235 & 0.0000 \\ \hline
        N=10 out=2 & 0.0570 & 3.3822 & 0.0004 \\ \hline
        N=10 out=3 & -0.4295 & -26.5330 & 0.0000 \\ \hline
        N=10 out=4 & -0.2229 & -7.8087 & 0.0000 \\ \hline
        N=10 out=5 & -0.2773 & -8.8486 & 0.0000 \\ \hline
        N=11 out=2 & 0.0630 & 3.9034 & 0.0000 \\ \hline
        N=11 out=3 & -0.4278 & -26.7720 & 0.0000 \\ \hline
        N=11 out=4 & -0.2716 & -10.6654 & 0.0000 \\ \hline
        N=11 out=5 & -0.2229 & -7.6797 & 0.0000 \\ \hline
        N=11 out=6 & -0.1746 & -5.3716 & 0.0000 \\ \hline
        N=12 out=2 & 0.0878 & 5.3122 & 0.0000 \\ \hline
        N=12 out=3 & -0.4619 & -28.5730 & 0.0000 \\ \hline
        N=12 out=4 & -0.2651 & -9.7207 & 0.0000 \\ \hline
        N=12 out=5 & -0.2729 & -9.6384 & 0.0000 \\ \hline
        N=12 out=6 & -0.0444 & -1.1916 & 0.2334 \\ \hline
    \end{tabular}
    \label{table1}
\end{table}

\subsubsection{Time-windowed Readout Controllers (dt Controllers)}

The trend between $e_{\Delta}(T)$ and the normed measures of \linebreak $\lVert s_{\Delta}(\xi_{\mu0},T) \rVert$ show a more complicated pattern than that of the t controllers, neither clearly conventional nor non-convention\-al. Rather, the overall trend shows a non-conventional positive correlation between $e_{\Delta}(T)$ and $\lVert s_{\Delta}(\xi_{\mu0},T) \rVert$ for target spins of $\ket{\OUT = 1}$ to $\ket{\OUT = 4}$ but a conventional, negative trend for transfers with $\ket{\OUT \geq 5}$. However, specifically for the tests concerning $e_{\Delta}(T)$ versus $\lVert s_{\Delta}(\xi_{\mu0},T) \rVert_{H}$ the test results in uniform refutation of $H_0$ in favor of $H_{1-}$ for the localization cases where $\ket{\OUT = 1}$ and with $p < \alpha = 0.01$ for all tests. Table~\ref{table2} provides a characteristic example of the Kendall $\tau$-based hypothesis test for $e_{\Delta}(T)$ versus $\lVert s_{\Delta}(\xi_{\mu0},T) \rVert_{H}$ for $N=3$ though $N=12$. In summary of the Kendall $\tau$-based hypothesis test for the time-windowed controllers we observe the following: 

\begin{itemize}

\item Of the $108$ test cases for the trend in $e_{\Delta}(T)$ versus \linebreak $\lVert s_{\Delta}(\xi_{\mu0},T) \rVert$, $21$ fail to meet the minimum confidence level and are not considered. However, for the $66$ cases of localization ($\ket{\OUT = 1}$) or transfers to $\ket{\OUT \leq 4}$, only three fail to meet the required confidence level. Of the remaining $63$ tests for localization or transfer up to $\ket{\OUT = 4}$, the hypothesis test rejects $H_0$ in favor of $H_{1+}$, a non-conven\-tional trend. In contrast, of the $42$ tests for transfer to $\ket{\OUT \geq 5}$, $18$ fail to meet the required confidence level. However, the remaining $24$ tests all display a negative, conventional trend, for these transfers.

\item For the tests of $e_{\Delta}(T)$ versus $\lVert s_{\Delta}(\xi_{\mu0},T) \rVert_{C}$, we see a higher percentage of tests that fail to meet the minimum confidence level, $36$ of $108$. In terms of trends, all localization or nearest-neighbor transfers show a non-conventional trend for sensitivity to controller uncertainty. Of the $56$ tests for transfers to $\ket{\OUT \geq 4}$, $24$ fail to make the cut, but the remaining $32$ test all show a conventional trend. Finally, we note that of the $16$ next-nearest-neighbor transfers, $14$ do not show a $p < 0.01$, and the two that do, for $N=5$ and $N=6$ display the non-conventional behavior.

\item The relation between $e_{\Delta}(T)$ and $\lVert s(\xi_{\mu0},T) \rVert_{H}$ shows a solid trend of conventional behavior for localization with a non-conventional trend for transfer to spins $\ket{\OUT \leq 4}$, but inconclusive results for the remaining cases. Specifically of the $18$ localization tests, all show a conventional trend with high confidence. Conversely, of the $48$ cases of transfer for $\ket{2 \leq \OUT \leq 4}$, all display a positive, non-conventional trend with $p < 0.01$. However, the remaining $42$ test cases fail to display a clear trend with the majority, $32$, failing to meet the required confidence level and the remainder displaying no clear trend.

\end{itemize}

\begin{table}[!t]
    \centering
    \caption{Excerpt of hypothesis test results for $e_{\Delta}(T)$ versus $\lVert s_{\Delta}(\xi_{\mu0},T) \rVert_{H}$ using the Kendall $\tau$. Of note are the conventional trends for all localization cases with $p < \alpha = 0.01$, and the non-conventional, positive trend for all transfer $\ket{2 \leq \OUT \leq 5}$ with strong confidence. The trend for transfer to $\ket{\OUT \geq 5}$ is inconclusive.}
    \begin{tabular}{|c|c|c|c|}
    \hline
        Transfer & \multicolumn{1}{p{2.5cm}}{\centering $\tau$ for $e_{\Delta}(T)$ vs. $\lVert s_{\Delta}(\xi_{\mu0},T) \rVert_{H}$}  & $Z_{\tau}$ & $p$ \\\hline
        N=3 out=1  & -0.6364 & -42.5924 & 0.0000 \\ \hline
        N=3 out=2  & 0.4810 & 32.1507 & 0.0000 \\ \hline
        N=4 out=1  & -0.3415 & -18.9007 & 0.0000 \\ \hline
        N=4 out=2  & 0.3509 & 13.3935 & 0.0000 \\ \hline
        N=5 out=1  & -0.6548 & -43.7457 & 0.0000 \\ \hline
        N=5 out=2  & 0.4997 & 30.1915 & 0.0000 \\ \hline
        N=5 out=3  & 0.4533 & 27.2474 & 0.0000 \\ \hline
        N=6 out=1  & -0.6431 & -35.4519 & 0.0000 \\ \hline
        N=6 out=2  & 0.2621 & 12.9521 & 0.0000 \\ \hline
        N=6 out=3  & 0.3070 & 18.0456 & 0.0000 \\ \hline
        N=7 out=1  & -0.6773 & -45.2345 & 0.0000 \\ \hline
        N=7 out=2  & 0.3447 & 18.5525 & 0.0000 \\ \hline
        N=7 out=3  & 0.2993 & 17.4544 & 0.0000 \\ \hline
        N=7 out=4  & 0.2610 & 12.5108 & 0.0000 \\ \hline
        N=8 out=1  & -0.7214 & -40.1470 & 0.0000 \\ \hline
        N=8 out=2  & 0.2749 & 15.5228 & 0.0000 \\ \hline
        N=8 out=3  & 0.3501 & 20.0874 & 0.0000 \\ \hline
        N=8 out=4  & 0.1286 & 4.6277 & 0.0000 \\ \hline
        N=9 out=1  & -0.7302 & -48.8551 & 0.0000 \\ \hline
        N=9 out=2 & 0.3203 & 16.8054 & 0.0000 \\ \hline
        N=9 out=3  & 0.3454 & 19.9429 & 0.0000 \\ \hline
        N=9 out=4  & 0.1102 & 4.4269 & 0.0000 \\ \hline
        N=9 out=5  & -0.0020 & -0.0718 & 0.4717 \\ \hline
        N=10 out=1  & -0.8023 & -44.2737 & 0.0000 \\ \hline
        N=10 out=2  & 0.2107 & 10.3089 & 0.0000 \\ \hline
        N=10 out=3  & 0.4274 & 25.0752 & 0.0000 \\ \hline
        N=10 out=4  & 0.1184 & 4.5067 & 0.0000 \\ \hline
        N=10 out=5  & 0.0448 & 1.4231 & 0.0773 \\ \hline
        N=11 out=1  & -0.7652 & -51.1965 & 0.0000 \\ \hline
        N=11 out=2  & 0.2871 & 15.0261 & 0.0000 \\ \hline
        N=11 out=3  & 0.4060 & 23.6861 & 0.0000 \\ \hline
        N=11 out=4  & 0.1651 & 7.3577 & 0.0000 \\ \hline
        N=11 out=5  & -0.0556 & -1.6805 & 0.0464 \\ \hline
        N=11 out=6  & 0.1131 & 3.2609 & 0.0006 \\ \hline
        N=12 out=1  & -0.8222 & -44.9496 & 0.0000 \\ \hline
        N=12 out=2  & 0.1577 & 7.9024 & 0.0000 \\ \hline
        N=12 out=3  & 0.4148 & 23.9986 & 0.0000 \\ \hline
        N=12 out=4  & 0.1737 & 7.6659 & 0.0000 \\ \hline
        N=12 out=5  & -0.0303 & -0.9387 & 0.1739 \\ \hline
        N=12 out=6  & -0.0571 & -1.3528 & 0.0881 \\ \hline
    \end{tabular}
    \label{table2}
\end{table}

As a check on consistency, we compare the Kendall $\tau$-based hypothesis test results with that obtained from the Pearson $r$. As with the case of the instant-readout controllers, we see strong agreement between the two measures:

\begin{itemize}

\item For $e_{\Delta}(T)$ versus $\lVert s_{\Delta}(\xi_{\mu0},T) \rVert$, the $66$ test cases for localization through $\ket{\OUT \leq 4}$, disagree in only three cases. The Kendall $\tau$ provides inconclusive results for $N = 6,\; 1 \rightarrow 3$ transfer and $N = 8$ localization, while the Pearson $r$ results show non-conventional trends for these transfers but is inconclusive on the $N=7,\; 1 \rightarrow 4$ transfer. In the remaining $63$ test cases for $\ket{\OUT \leq 4}$, the tests agree on a non-conventional trend. While of the remaining $42$ test cases, the Pearson $r$ results in $21$ inconclusive tests versus $20$ for the Kendall $\tau$, all cases in which both tests present $p < 0.01$ agree on a conventional trend for these transfers.

\item For controller uncertainty, the $e_{\Delta}(T)$ versus $\lVert s(\xi_{\mu0},T) \rVert_{C}$ trends show perfect agreement in rejecting $H_0$ in favor of $H_{1+}$ for all localization and nearest-neighbor transfers. In terms of the next-nearest-neighbor transfers (those transfers to $\ket{OUT = 3}$), the Pearson-based test agrees with the Kendall $\tau$-based test in rejection of $H_{0}$ in favor of $H_{1+}$ for $N=5$ and $N=6$. However, for the remaining $14$ next-nearest-neighbor transfer, the Pearson $r$ test statistic provides inconclusive results. For the $56$ test cases for $\ket{OUT \geq 4}$, the Pearson $r$-based test returns $18$ instances that fall below the confidence threshold. However, in all cases where both the Kendall $\tau$ and Pearson $r$ present high confidence, the hypothesis test agrees in rejection of $H_0$ for $H_{1-}$ for these transfers.

\item Of the $108$ test cases for $e_{\Delta}(T)$ versus $\lVert s(\xi_{\mu0},T) \rVert_{H}$, we see agreement between both measures in $100$ cases. The $8$ conflicts arise from one test or the other failing to reject the null hypothesis while the other does reject $H_0$, but in no cases to both tests reject $H_0$ in favor of opposing alternative hypotheses. Of note, for the conventional trend of localization assessed by the Kendall $\tau$-based test, the Pearson $r$ test agrees on all counts save for $N=5$ and $N=12$, which are inconclusive based on the Pearson $r$.

\end{itemize} 

\subsection{Equivalent Error---Widely Varying Robustness}

\begin{figure}[!t]
\centering
\includegraphics[width=1\textwidth]{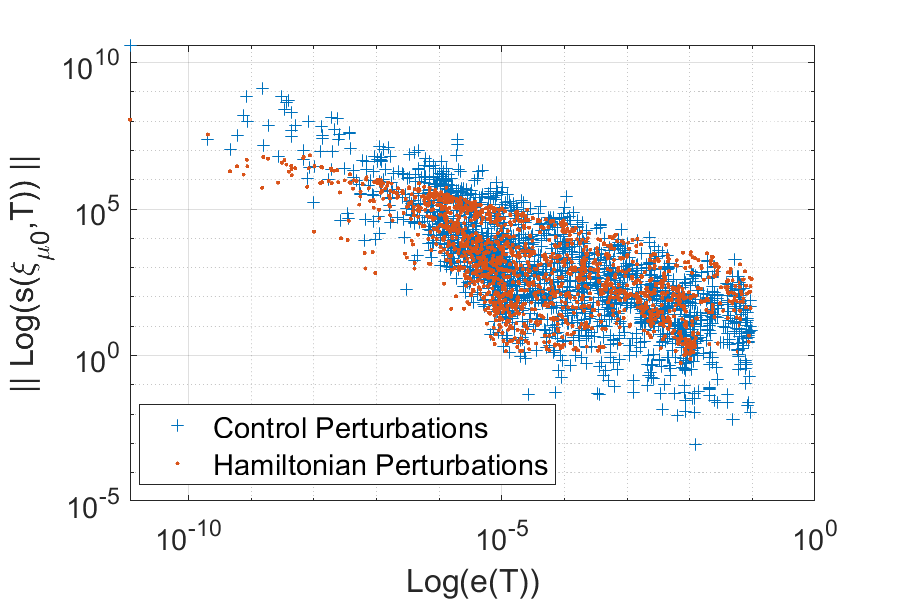}
\caption{Log-log plot of $\lVert s(\xi_{\mu0},T) \rVert_{C}$ (blue crosses) and $\lVert s(\xi_{\mu0},T) \rVert_{H}$ (red dots) versus $e(t)$ for a nearest-neighbor transfer in a $5$-ring. Note the overall negative (conventional) trend, but also the variation in log-sensitivity by orders of magnitude for controllers on the same vertical line.}
\label{N=5 1-2t}
\end{figure}

\begin{figure}[!t]
\centering
\includegraphics[width=1\textwidth]{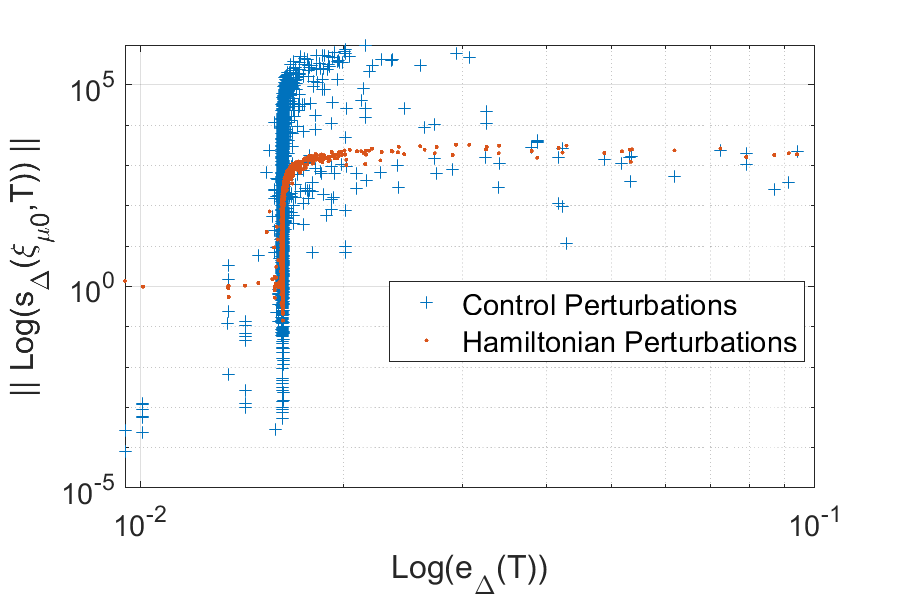}
\caption{Log-log plot of $\lVert s_{\Delta}(\xi_{\mu0},T) \rVert_{C}$ (blue crosses) and $\lVert s_{\Delta}(\xi_{\mu0},T) \rVert_{H}$ (red dots) versus $e_{\Delta}(t)$ for a nearest-neighbor transfer in a $3$-ring for time-windowed readout. Note the major variations in log-sensitivity for controllers with an error in the range of $0.016$.}
\label{N=3 1-2dt}
\end{figure}

Though the hypothesis test of Section~\ref{results} provides insight into the trends of error versus log-sensitivity on a large scale, it does not tell the entire story. In fact, one of the more interesting features of the controllers in this data set is the range of log-sensitivity observed for a given fidelity error. Figure~\ref{N=5 1-2t} displays the log-sensitivity for controller and Hamiltonian perturbations versus error for instant-time readout (t controllers) in a $5$-ring with nearest-neighbor transfer. The overall trend of the figure confirms the negative trend of the hypothesis test, but the spread of log-sensitivities for a given error is large.  For example, the log-sensitivity for controllers with a fidelity error $e(T) = 10^{-5}$ ranges from as low as $10^0$ to greater than $10^5$. This belies a simple one-parameter relation between log-sensitivity and error, but provides evidence for the existence of controllers with the best of possible properties: good performance and with acceptable robustness. As a second example, we show the plot of $\lVert s_{\Delta}(\xi_{\mu0},T) \rVert_C$ versus $e_{\Delta}(T)$ for a $3$-ring for nearest-neighbor transfer and time-windowed readout (dt controller) in Figure~\ref{N=3 1-2dt}. The plot confirms the positive (non-conventional) trend of the hypothesis test but displays wide variation in log-sensitivities in the vicinity of $e_{\Delta}(T) = 0.016$ from as low as $10^{-3}$ upwards to $10^{5}$. Identification of what factors guarantee the smaller log-sensitivities or prevent the larger values would be highly beneficial in the process of controller design and selection, but remain an open question.

\begin{figure}[!t]
\centering
\includegraphics[width=1\textwidth]{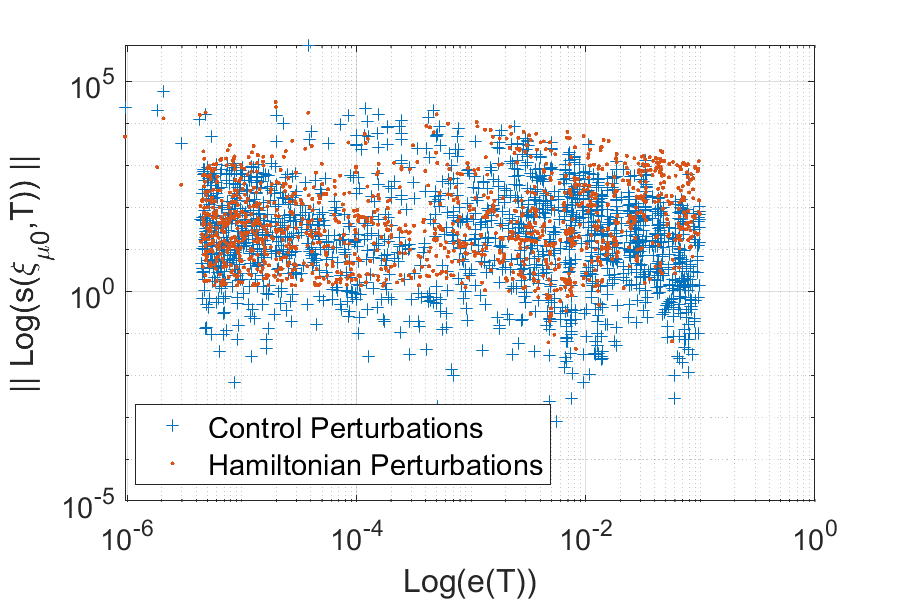}
\caption{Log-log plot of $\lVert s(\xi_{\mu0},T) \rVert_{C}$ (blue crosses) and $\lVert s(\xi_{\mu0},T) \rVert_{H}$ (red dots) versus $e(t)$ for a nearest-neighbor transfer in a $7$-ring and instantaneous readout. Note that a strong positive trend is not visually apparent from the plot, but the plot does display the same characteristic of widely varying log-sensitivities for the same error, suggesting the ability to select controllers with good robustness and performance.}
\label{N=7 1-2t}
\end{figure}

Next, we note the visual depiction of the nearest-neighbor transfers for instantaneous readout controllers with $N\geq7$ in Figure~\ref{N=7 1-2t}. Though the hypothesis test results show rejection of $H_0$ in favor of $H_{1+}$ for these cases, the trend is not readily apparent visually as seen in Figure~\ref{N=7 1-2t}. This can be confirmed by the relatively small values of the Kendall $\tau$ and Pearson $r$ for these transfers. However, of greater importance are the variations in the log-sensitivity for a given error seen in the plot, again indicating the possibility of controllers that provide good robustness for acceptable performance.

\begin{figure}[!t]
\centering
\includegraphics[width=1\textwidth]{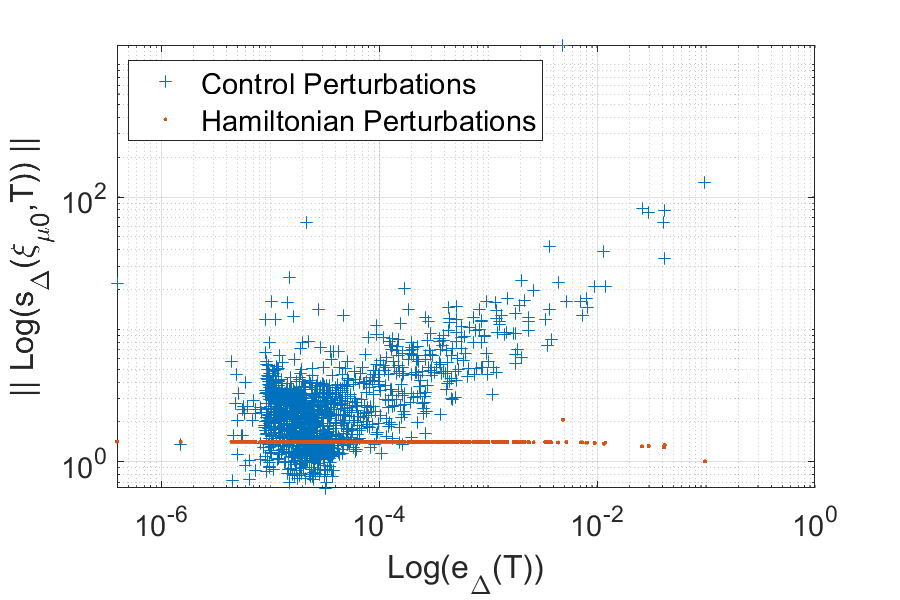}
\caption{Log-log plot of $\lVert s_{\Delta}(\xi_{\mu0},T) \rVert_{C}$ (blue crosses) and $\lVert s_{\Delta}(\xi_{\mu0},T) \rVert_{H}$ (red dots) versus $e_{\Delta}(t)$ for $6$-ring localization. Note the negative trend for controller uncertainty but almost flat trend for Hamiltonian uncertainty.}
\label{N=6 1-1dt}
\end{figure}

Finally, we look at the plot of a localization case in Figure~\ref{N=6 1-1dt}. We clearly see the contrast in robustness for localization between Hamiltonian uncertainty and controller uncertainty. The strong non-conventional trend between $e_{\Delta}(T)$ and $\lVert s_{\Delta}(\xi_{\mu0},T) \rVert_{C}$ is clearly evident while the slightly negative conventional trend for $\lVert s_{\Delta}(\xi_{\mu0},T) \rVert_{H}$ is also perceptible. But more important is the nearly constant value of the log-sensitivity for Hamiltonian uncertainty over the range of error, a factor that can likely be exploited to provide some robustness guarantees over large performance ranges in the case of localization.  

\section{Conclusion} \label{conclude}

In this paper we use a basic hypothesis test to determine the degree by which controllers optimized for coherent excitation transport in quantum rings abide by the limitations implied by classical control, extending the work initiated in \textcite{Jonckheere2018}. In contrast to \textcite{Jonckheere2018}, we extended the analysis to consider not only controllers optimized for time-averaged fidelity, but those optimized for instantaneous readout as well. Furthermore, we included uncertainty in both the controlling bias fields and the spin-couplings. Overall, our results confirm those of \textcite{Jonckheere2018} in that controllers optimized for readout over a time window exhibit a degree of non-classical behavior for transfer to spins near the initial spin and regain conventional behavior for transfers between more distant spins. However, while the results by \textcite{Jonckheere2018} indicate non-conventional trends for the localization cases with Hamiltonian perturbations, using the updated calculations of~\eqref{eq:sens_dt} yields more conventional results based on the Kendall $\tau$ and Pearson $r$ hypothesis tests. In the extension of the analysis to controllers optimized for instantaneous readout, we note a strong conventional trend for all spin sizes and transfers, save for the nearest-neighbor transfers of $N\geq7$. Finally, we show that beyond just the hypothesis testing, controllers of both types display widely varying levels of robustness for the same error.

Looking to future work, we need to identify what drives the variation in log-sensitivity for controllers with similar error in order to direct synthesis towards controllers that provide the best robustness properties for a given fidelity requirement. Next, the cause for the differences in the log-sensitivity trends observed for controllers optimized for instantaneous readout versus readout over a time window, and transfer to nearest-neighbor and next-nearest-neighbor spins in both types of controllers needs to be clarified. This holds the potential to exploit these properties to navigate around the classically imposed fundamental limitations. Finally, it is necessary to generalize the one-uncertainty-at-a-time nature of the differential sensitivity technique used in this paper to more general methods that account for multiple structured uncertainties or even unstructured uncertainties.

\section*{Conflict of interest}

The authors report no conflicts of interest.

\section*{Financial support}

Sean O'Neil acknowledges PhD funding from the US Army Advanced Civil Schooling program.

\section*{Data availability}

The data is available at \textcite{DataSet1}.

\printbibliography

\end{document}